\documentclass[page-classic]{epl2} 

\usepackage{geometry}                		
\usepackage{graphicx}				
\usepackage{amssymb,amsmath}

\def\0{\mbox{\boldmath$\displaystyle\mathbb{O}$}}

\def\s{\mbox{\boldmath$\displaystyle\boldsymbol{\sigma}$}}
\def\p{\mbox{\boldmath$\displaystyle\boldsymbol{p}$}}

\def\beq{\begin{equation}}
\def\eeq{\end{equation}}

\def\bea{\begin{align}}
\def\ena{\end{align}}

\title{Elko under spatial rotations}

\author{Dharam Vir Ahluwalia\inst{1} and Sweta Sarmah\inst{2}}

\institute{                    
  \inst{1} Centre for the Studies of the Glass Bead Game Chaugon, Bir, Himachal Pradesh 176 077, India\\
  \inst{2} Department of Biochemistry and  Tissue Biology, State University of Campinas, S\~ao Paulo 13083-859, Brazil.
}

\pacs{04.20.Gz}{Spacetime topology, causal structure, spinor structure}
\pacs{03.65.Vf}{Formalism}
\pacs{03.65.Vf}{Phases: geometric; dynamic or topological}

\abstract{
Under a
rotation by an angle $\vartheta$, both the right- and left- handed Weyl spinors pick up a phase factor ${\exp(\pm\, i \vartheta/2)}$. 
The upper sign holds for the positive helicity spinors, while the lower sign for the negative helicity spinors.
For $\vartheta = 2\pi$ radians this produces the famous minus sign.
However, the four-component spinors are built from a direct sum of the indicated two-component spinors. The effect of the rotation by $2\pi$ radians on the eigenspinors of the parity -- that is, the Dirac spinors -- is the same as on Weyl spinors.
It is because for these spinors the right- and left- transforming components have the same helicity~\cite{Weinberg:1995mt,Ahluwalia:2016jwz}. And the rotation induced phases, being same, factor out.
  But for the eigenspinors of the charge conjugation  operator, i.e. Elko,  the left- and right- transforming components have opposite helicities, and therefore they pick up opposite phases. As a consequence the 
  behaviour of the eigenspinors of the charge conjugation operator (Elko) is more subtle: for $0<\vartheta<2\pi$ a self conjugate spinor becomes a linear combination of the self and antiself conjugate spinors with $\vartheta$ dependent superposition coefficients -- and yet the rotation preserves the self/antiself conjugacy  of these spinors! This apparently paradoxical situation is fully resolved.
This new effect, to the best of our knowledge, has never been reported before. The purpose of this communication is to present this result and to correct an interpretational error of a previous version. }

\begin{document}
\maketitle

The eigenspinors of the charge conjugation operator known by their German acro\-nym Elko, and also as dark spinors, serve as expansion coefficients of mass dimension one fermionic fields of spin one half~\cite{Ahluwalia:2004sz,Ahluwalia:2004ab,Ahluwalia:2016rwl,Ahluwalia:2016jwz,HoffdaSilva:2016ffx,daRocha:2005ti}.
They have been extensively studied in 
cosmology, see for example~\cite{Boehmer:2010ma,Fabbri:2010ws,Pereira:2017bvq,Pereira:2017efk,Pereira:2016emd} 
 and~\cite[Section 7.1]{Bahamonde:2017ize}. For more than a decade it is known that the right- and left- transforming components of Elko
necessarily have opposite helicities and that they are not annihilated by the Dirac operator $\left(\gamma^\mu p_\mu \pm m \right)$. The former property has the further consequence that we report here, while the latter leads to mass dimension one fermions when Elko are used as an expansion coefficients of a spin one half quantum field. The new fermions also serve as first principle dark matter candidates.

To obtain the abstracted result, we need eigenspinors of the charge conjugation operator, that is, Elko.
A systematic presentation of these is given in~\cite{Ahluwalia:2016rwl}, with their differences from the Majorana field and Majorana spinors are summarised in the section that follows these introductory remarks.  Elko are denoted by $\lambda_\alpha^S(\p)$ and $\lambda_\alpha^A(\p)$. Under the charge conjugation operator, they have the  doubly degenerate eigenvalues $+1$ and $-1$ respectively. The degeneracy index $\alpha=\pm 1$ is related to the helicity of the Weyl spinors that enter the definition of these spinors. The symbol $\p$ refers to the linear momentum of the associated particles.

\section{Majorana field, Majorana spinors, and Elko}

In the usual notation, the 1937 Majorana field~\cite{Majorana:1937vz} sets $b^\dagger(\p,\sigma)$ equal to $a^\dagger(\p,\sigma)$ in the Dirac field $\psi(x)$. This results in an intrinsically neutral field, but it is still expanded in terms of Dirac spinors. The right transforming and left transforming components of this field are related as~\cite{Weinberg:1996kr}
\begin{equation}
\psi_R(x) = \Theta \psi^\ast_L(x)
\end{equation}
where $\Theta$ is the Wigner time reversal operator for spin one half~\cite{Ahluwalia:2016rwl}. For this reason the Majorana \textit{field} can be written as
\begin{equation}
\psi_M(x) = \left(\begin{array}{c}
\Theta \psi^\ast_L(x)\\
\psi_L(x)
\end{array}\right).\label{eq:Mf}
\end{equation}
On the other hand Majorana \textit{spinors}, $\lambda(\p)$, have the form\footnote{The superficial similarity between $\psi_M(x)$ and $\lambda(\p)$  is sometimes mistakenly taken as to mean that Majorana spinors define Majorana field.}
\begin{equation}
\lambda(\p) = \left(\begin{array}{c}
\Theta \phi^\ast(\p)\\\
\phi(\p)
\end{array}\right)\label{eq:S}
\end{equation}
where $\phi(\p)$ is a two-component left-transforming Weyl spinor. Under the charge conjugation operator operator, $\mathcal{C}$, it is self-conjugate, that is 
\begin{equation}
\mathcal{C} \lambda(\p) = + \lambda(\p).
\end{equation}
In the formalism of mass dimension one fermions it was realised that the literature had missed the \textit{negative} eigenvalue, and a \textit{complete set} of 
eigenspinors  of $\mathcal{C}$ were introduced
\begin{equation}
\mathcal{C} \lambda^S(\p) = + \lambda^S(\p),\quad
\mathcal{C} \lambda^A(\p) = - \lambda^A(\p).
\end{equation}
The complete set of spinors are known by their German acronym Elko.

An attempt to construct a Lagrangian density for Majorana spinors by the authors of~\cite{Aitchison:2004cs} failed and they argued, as many authors before, that Majorana spinors should be treated as Grassmann numbers. The development of Elko showed that it was not necessary, and that Elko can be treated at par with Dirac spinors. In the meantime it had remained unappreciated that,  (a)  under the Dirac dual Majorana spinors have a \textit{null} norm, (b) that they are \textit{not} annihilated by the Dirac operator
$(\gamma_\mu p^\mu \mp m\mathbb{I})$, and (c) the right- and left- transforming components of Majorana spinors are endowed with \textit{opposite} helicities. All this, and many other properties of Majorana spinors, came to be known in the development of Elko. 

To avoid confusion as to which eigenspinors of the charge conjugation operator one was talking about it was deemed necessary that a new name be introduced, and Elko thus came to complete the story of Majorana spinors (but not in their Grassmann incarnation).

It is also worth noting that  Majorana spinors were always considered as four-component spinors disguising the two-component Weyl spinors, Elko formalism showed that $\lambda(\p)$, when complemented by $\lambda^A(\p)$, annuls this folklore.


\section{Setting up the co-ordinate system}

Understanding the new spinors vastly simplifies in the basis we have chosen for them~\cite{Ahluwalia:2016jwz}. Their right- and left- transforming components have their spin projections assigned not with respect an external $\widehat{z}$-axis but to a self referential unit vector associated with their motion: that is $\widehat{\textbf{p}}$. We thus find it convenient to erect an orthonormal cartesian coordinate system with $\widehat{\textbf{p}}$ as one of its axis.
With $\widehat{\textbf{p}}$ chosen as\footnote{Where $\sin[\theta]$ is abbreviated as $s_\theta$ with obvious extension to other trigonometric function.} 
\begin{equation}
\left(s_\theta c_\phi, s_\theta s_\phi,c_\theta\right)
\end{equation}
 we introduce two more unit vectors 
 \begin{equation}
 \begin{split}
\widehat{\eta}_+ & \stackrel{\textrm{def}}{=}\frac{1}{\sqrt{2+a^2}}\left(1,1,a\right), \quad a \in \Re \\
\widehat{\eta}_-  &  \stackrel{\textrm{def}}{=} 
\frac{\widehat{\textbf{p}}\times\widehat{\eta}_+ }
{\sqrt{\left(\widehat{\textbf{p}}\times\widehat{\eta}_+ \right)
\cdot 
\left(\widehat{\textbf{p}}\times\widehat{\eta}_+ \right)}}
\label{eq:etam}
\end{split}
\end{equation}
and impose the requirement
\begin{equation}
\begin{split}
& \widehat{\eta}_+\cdot \widehat{\eta}_- = 0,\quad
\widehat{\eta}_+\cdot \widehat{\textbf{p}} = 0,\quad
\widehat{\eta}_-\cdot \widehat{\textbf{p}} = 0\\
& \widehat{\eta}_+\cdot \widehat{\eta}_+ = 1,\quad
\widehat{\eta}_-\cdot \widehat{\eta}_- = 1.
\end{split}
\end{equation}
Requiring $\widehat{\eta}_+\cdot \widehat{\textbf{p}} $ to vanish 
reduces 
$\widehat{\eta}_+ $ to
\begin{equation}
\widehat{\eta}_+ = \frac{1}{\sqrt{2 + (1+s_{2\phi})t^2_\theta}}\Big( 1,1,-(c_\phi + s_\phi)t_\theta\Big)
\end{equation}
Definition (\ref{eq:etam}) then immediately yields
\begin{equation}
\widehat{\eta}_- = \frac{1}{\sqrt{2 + (1+s_{2\phi})t^2_\theta}}\Big(  -c_\theta - s_\theta s_\phi(c_\phi+s_\phi)t_\theta, \,
 c_\theta + s_\theta c_\phi(c_\phi+s_\phi)t_\theta,\,
 s_\theta(c_\phi-s_\phi) \Big).
\end{equation}

In the limit when both the $\theta$ and $\phi$ tend to zero, the above-defined unit vectors take the form
\begin{equation}
\widehat{\eta}_+\big\vert_{\theta\to 0,\phi\to 0} = \frac{1}{\sqrt{2}}\left(1,1,0\right), \quad
\widehat{\eta}_-\big\vert_{\theta\to 0,\phi\to 0} = \frac{1}{\sqrt{2}}\left(-1,1,0\right)\label{eq:etaminus}
\end{equation}
and orthonormal system $\left(\widehat{\eta}_-,\widehat{\eta}_+,\widehat{\textbf{p}}\right)$
 does not reduce to the standard cartesian coordinate system  
$(\widehat{\textbf{x}},\widehat{\textbf{y}},\widehat{\textbf{z}})$. For this reason, without destroying the orthonormality of the introduced unit vectors and guided by  (\ref{eq:etaminus}), 
 we exploit the freedom of a rotation about the $ \widehat{\textbf{p}}$ axis (in the plane defined by $\widehat{\eta}_+$ and $\widehat{\eta}_-$) and introduce
 \begin{equation}
 \widehat{\textbf{p}}_\pm \stackrel{\textrm{def}}{=} \frac{1}{\sqrt{2}}\left(\widehat{\eta}_+ \pm \widehat{\eta}_-\right).
 \end{equation}
The set of axes $\left(\widehat{\textbf{p}}_-,\widehat{\textbf{p}}_+,\widehat{\textbf{p}}\right)$ do indeed form a right-handed coordinate system that reduces to the standard cartesian system in the limit  both the $\theta$ and the $\phi$ tend to zero.

 
\section{Establishing the spin of Elko}
We now introduce three generators of rotations about each of the three axes
\begin{equation}
\texttt{J}_- \stackrel{\textrm{def}}{=} \frac{\s}{2}\cdot \widehat{\textbf{p}}_-
,\quad 
\texttt{J}_+ \stackrel{\textrm{def}}{=} \frac{\s}{2}\cdot \widehat{\textbf{p}}_+
,\quad 
\texttt{J}_p \stackrel{\textrm{def}}{=} \frac{\s}{2}\cdot \widehat{\textbf{p}}
\end{equation}
$\texttt{J}_p $ coincides with the helicity operator $\mathfrak{h} $. As a check, a straightforward exercise shows that the three $\texttt{J}'s$ satisfy 
the $\mathfrak{s}\mathfrak{u}(2)$ algebra needed for the generators of rotation
\begin{equation}
\left[\texttt{J}_-,\texttt{J}_+\right] = i  \texttt{J}_p,\quad
\left[\texttt{J}_p,\texttt{J}_-\right] = i  \texttt{J}_+,\quad
\left[\texttt{J}_+,\texttt{J}_p\right] = i  \texttt{J}_-.
\end{equation}
Since the eigenspinors of the charge conjugation operator reside in the $\mathcal{R}\oplus\mathcal{L}$ representation space --
with $\mathcal{R}$ and $\mathcal{L}$ standing for spin one half  Weyl spaces of the right and left type --
 and as far as rotations are concerned both the $\mathcal{R}$ and
$\mathcal{L}$ spaces are served by the same generators of rotations. We therefore
introduce
\begin{equation}
\texttt{h}_ - =\left(\begin{array}{cc}
{\mathtt{J}_-} & \texttt{0}_2\\
\texttt{0}_2 &{\mathtt{J}_-} 
\end{array}
\right),\quad
\texttt{h}_ + =\left(\begin{array}{cc}
{\mathtt{J}_+} & \texttt{0}_2\\
\texttt{0}_2 &{\mathtt{J}_+} 
\end{array}
\right),\quad
\texttt{h}_ p =\left(\begin{array}{cc}
{\mathtt{J}_p} & \texttt{0}_2\\
\texttt{0}_2 &{\mathtt{J}_p} 
\end{array}
\right)
\end{equation}
where $\texttt{0}_2$ is a $2\times 2$ null matrix. A straight forward calculation then yields the result
\begin{eqnarray}
\textrm{h}_p \,\lambda^S_+(\p) &= -\frac{1}{2}\lambda^A_-(\p),\quad
\textrm{h}_p \,\lambda^S_-(\p) = -\frac{1}{2}\lambda^A_+(\p)\label{eq:hplamdaS}\\
\textrm{h}_p \,\lambda^A_+(\p) & = -\frac{1}{2}\lambda^S_-(\p),\quad
\textrm{h}_p \,\lambda^A_-(\p)  = -\frac{1}{2}\lambda^S_+(\p)
\label{eq:hplamdaA}
\end{eqnarray}
with the consequence that each of the $\lambda(\p)$
 is an eigenspinor of $\texttt{h}_p^2$
\begin{equation}
\textrm{h}_p^2 \,\lambda^{S,A}_\pm(\p) = \frac{1}{4}\lambda^{S,A}_\pm(\p).
\end{equation}
The action of $\texttt{h}_-$ and $\texttt{h}_+$ is more involved,  
for instance
\begin{equation}
\texttt{h}_+\,\lambda^S_+(\p) =  - \frac{1}{2}\left(
\alpha \lambda^S_-(\p) + \beta\lambda^A_+(\p)\right)
\end{equation}
with
\begin{eqnarray}
\alpha =  - \frac{im (m+E)\big(\cos\phi(1+\sec\theta)+(-1+\sec\theta)\sin\phi\big)}
{\sqrt{4+2(1+\sin 2\phi)\tan^2\theta}}\\
\beta=  - \frac{m (m+E)\big(\cos\phi(-1+\sec\theta)+(1+\sec\theta)\sin\phi\big)}
{\sqrt{4+2(1+\sin 2\phi)\tan^2\theta}}
\end{eqnarray}
but the action of their squares is much simpler and reads
\begin{equation}
\textrm{h}_-^2 \,\lambda^{S,A}_\pm(\p) = \frac{1}{4}\lambda^{S,A}_\pm(\p),\quad
\textrm{h}_+^2 \,\lambda^{S,A}_\pm(\p) = \frac{1}{4}
\lambda^{S,A}_\pm(\p).
\end{equation}
This exercise thus establishes that 
\begin{equation}
\texttt{h}^2 \stackrel{\textrm{def}}{=} \textrm{h}_- ^2+ \textrm{h}_+^2 + \textrm{h}_p^2
\end{equation}
while acting on each of the $\lambda(\p)$ yields 
\begin{equation}
\texttt{h}^2 \lambda^{S,A}_\pm = \frac{3}{4}  \lambda^{S,A}_\pm =\frac{1}{2}
\left(1+ \frac{1}{2}\right) \lambda^{S,A}_\pm
\end{equation}
and confirms spin one half for $\lambda(\p)$.

For ready reference, we note the counterpart of (\ref{eq:hplamdaS}) and 
 (\ref{eq:hplamdaA}) for the Dirac spinors: 
 \begin{eqnarray}
\textrm{h}_p \, u_+(\p) &= \frac{1}{2} u_+(\p),\quad
\textrm{h}_p \,u_-(\p)  = -\frac{1}{2}u_-(\p) 
\label{eq:hplamdau}\\
\textrm{h}_p \,v_+(\p) & = -\frac{1}{2}v_+(\p),\quad
\textrm{h}_p \,v_-(\p)  = \frac{1}{2}v_-(\p).\label{eq:hplamdav}
\end{eqnarray}

\section{The result}
For simplicity we consider a rotation by 
$\vartheta$ about $\widehat{\p}$ axis and find that  a $2\pi$ rotation induces the expected minus sign for $\lambda(\p)$, but for a general rotation it mixes the self and antiself conjugate spinors:
\begin{eqnarray}
\exp\left({i \texttt{h}_p \vartheta}\right)\, \lambda^S_\pm(\p) = \cos\left({\vartheta}/{2}\right) \lambda^S_\pm(\p) - i \sin\left({\vartheta}/{2}\right) \lambda^A_\mp(\p)
\label{eq:22}\\
\exp\left({i \texttt{h}_p \vartheta}\right)\, \lambda^A_\pm(\p) = \cos\left({\vartheta}/{2}\right) \lambda^A_\pm(\p) - i \sin\left({\vartheta}/{2}\right) \lambda^S_\mp(\p).
\label{eq:23}
\end{eqnarray}
In contrast, for the Dirac spinors we have the well-known result
\begin{equation}
\exp\left({i \texttt{h}_p \vartheta}\right) \psi_\pm(\p) = \exp(\pm i\vartheta/2) \psi_\pm(\p) 
\end{equation}
where $\psi_\pm(\p)$ stands for any one of the four $u_\pm(\p)$ and $v_\pm(\p)$ spinors of Dirac. 

\texttt{Do the results (\ref{eq:22}) and (\ref{eq:23}) imply that rotation induces loss of self/anti-self\\ conjugacy under  charge conjugation $\mathcal{C}$?} The answer is: no. To see this we observe that
\begin{eqnarray}
\mathcal{C} \left[ i \lambda^A_{\pm}(\p)\right] &=& (-i)(-\lambda^A_{\pm}(\p))
=+ \left[ i \lambda^A_{\pm}(\p)\right] \label{eq:obs1}\\
\mathcal{C} \left[ i \lambda^S_{\pm}(\p)\right] &=& (-i)(\lambda^S_{\pm}(\p))
=- \left[ i \lambda^S_{\pm}(\p)\right].\label{eq:obs2}
\end{eqnarray}
We thus define a set of new self and anti-self conjugate spinors (see (\ref{eq:22}) and (\ref{eq:23}))
\begin{eqnarray}
 \lambda^s(\p) &\stackrel{\text{def}}{=}& \cos\left({\vartheta}/{2}\right) \lambda^S_\pm(\p) - i \sin\left({\vartheta}/{2}\right) \lambda^A_\mp(\p) \\
 \lambda^a(\p)& \stackrel{\text{def}}{=} &\cos\left({\vartheta}/{2}\right) \lambda^A_\pm(\p) - i \sin\left({\vartheta}/{2}\right) \lambda^S_\mp(\p)
\end{eqnarray}
and verify that 
 \begin{equation}\mathcal{C}\lambda^{s}_\pm (\p) = + \lambda^{s}_\pm (\p), \quad\mathcal{C}\lambda^{a}_\pm (\p) = - \lambda^{a}_\pm (\p).
\end{equation} 
As a result (\ref{eq:22}) and  (\ref{eq:23}) reduce to
\begin{eqnarray}
\exp\left({i \texttt{h}_p \vartheta}\right)\, \lambda^S_\pm(\p) = \lambda^s_\pm(\p) \\
\exp\left({i \texttt{h}_p \vartheta}\right)\, \lambda^A_\pm(\p) = \lambda^a_\pm(\p)
\end{eqnarray}
and confirm that rotation preserves $\mathcal{C}$-self/anti-self conjugacy of Elko. The explicit expressions for the new set of Elko are now readily obtained, and read:
\begin{equation}
\lambda^s_\pm(\p) = \varrho_\pm\lambda^S_\pm(\p),\quad
\lambda^a_\pm(\p) = \varrho_\mp\lambda^A_\pm(\p)
\end{equation}
where the $4\times 4$ matrices $\varrho_\pm$ are defined as
\begin{equation}
\varrho_\pm =\left(
\begin{array}{cc}
e^{\mp i\vartheta/2} \mathbb{I} & \mathbb{O}\\
\mathbb{O} & e^{\pm i\vartheta/2} \mathbb{I}
\end{array}
\right).
\end{equation}
In the above expression, $\mathbb{I}$ and $\mathbb{O}$ are $2\times2$ identity and null matrices respectively. Apart from the special 
values of $\vartheta$ for which
\begin{equation}
\varrho_\pm =\bigg\{\begin{array}{ll}
- 1, & \mbox{for $\theta = 2\pi$}\\
+ 1,& \mbox{for $\theta = 4\pi$}.
\end{array}
\end{equation}
 Elko and Dirac spinors behave differently.
 
 The next question thus arises: Since in all physical observables  Elko appear as bilinears, do the sets $\{\lambda^S_\pm(\p),
 \lambda^A_\pm(\p)\}$ and  $\{\lambda^s_\pm(\p),
 \lambda^a_\pm(\p)\}$ yield identical results? The answer is: yes. It comes about because
 \begin{equation}
 (\varrho_\pm)^\dagger \gamma_0 \varrho_\mp = \gamma_0.
 \end{equation}
 So from the orthonormality relations, to the completeness relations, to the spin sums, the two sets 
 $\{\lambda^S_\pm(\p),
 \lambda^A_\pm(\p)\}$ and  $\{\lambda^s_\pm(\p),
 \lambda^a_\pm(\p)\}$  carry identical results.


\section{Concluding remarks}

Both the  Dirac spinors and Elko arise as representations of the extended Lorentz algebra. The former are eigenspinors of the parity operator, while the latter arise as eigenspinors of the charge conjugation operator. This difference makes left- and right- transforming components of Dirac spinors carry \textit{same} helicities, while for Elko it yields \textit{opposite} helicities. The exception occurs when the rotation angle $\vartheta$ equals $2\pi$ or $4\pi$.  The emergence of minus under $2\pi$ rotation has given rise to lively discussion and 
 ingenious experiments~\cite{Aharonov:1967zz,Dowker:1969ia,RAUCH1974369,Werner:1975wf,Silverman:1980mp,Horvathy:1984sm,Klein:1976qc}. But that is only for the Dirac spinors. This communication hopes to inspire a similar discussion for Elko. The unearthed theoretical structure underlying Elko suggests that further phenomenological and theoretical  implications must be investigated, particularly because Elko define mass dimension one fermions.

\acknowledgments We acknowledge discussions with the Guaratinguet\'a group working on Elko and mass dimension one fermions. One of us (DVA) also thanks Swagat Mishra and Thanu Padmanabhan  for asking good questions. 

\section{Addendum} In the first two arXived versions  of this communication we failed to note observations~(\ref{eq:obs1}) and (\ref{eq:obs2}), see reference~\cite{Ahluwalia:2018hfm}. It led to an erroneous conclusion. We correct it here. Specifically, the answer to the question asked just above (30) is \textit{not} a  yes, but a no.



\end{document}